\documentclass[twocolumn,prd,aps,nofootinbib,showpacs]{revtex4}
\usepackage{multirow}
\usepackage{graphicx}
\usepackage[dvips]{color}
\usepackage{amssymb,amsmath}

\begin{document}

\title{Rummukainen-Gottlieb's formula on two-particle system with different mass}

\author{Ziwen Fu}

\affiliation{
Key Laboratory of Radiation Physics and Technology {\rm (Sichuan University)},
Ministry of Education; \\
Institute of Nuclear Science and Technology, Sichuan University,
Chengdu 610064, P. R. China.
}

\begin{abstract}
L\"uscher established a non-perturbative formula to
extract the elastic scattering phases from
two-particle energy spectrum in a torus using lattice simulations.
Rummukainen and Gottlieb further extend it to the moving frame,
which is devoted to the system of two identical particles.
In this work, we generalize Rummukainen-Gottlieb's formula to
the generic two-particle system where two particles are explicitly distinguishable, namely, the masses of the two particles are different.
The finite size formula are achieved for both $C_{4v}$ and $C_{2v}$ symmetries.
Our analytical results will be very helpful for the study of some resonances,
such as kappa, vector kaon, and so on.
\end{abstract}

\pacs{ 12.38.Gc, 11.15.Ha }

\maketitle

\section{Introduction}
Many low energy hadrons, such as kappa, sigma,
can observed as resonances in the experiments.
The energy eigenvalues of two-particle systems
can be achieved by calculating the propagators using lattice QCD.
Hence, it is highly desired to connect
these calculated energy eigenstates
to the experimental scattering phases.
L\"uscher~\cite{Luscher:1986pf,Luscher:1991cf,
Lellouch:2001p4241,Luscher:1991ux,Luscher:1990ck} have established
a non-perturbative formula to connect
two-particle system's energy in a torus
with the elastic scattering phase.
Rummukainen and Gottlieb further extended L\"uscher's formula
to the moving frame (MF)~\cite{Rummukainen:1995vs}.
Moreover, Xu Feng et al generalized  L\"uscher's formula in an asymmetric box~\cite{Feng:2004ua}.
These formulae have been employed in many different
applications~\cite{Fu:2011bz,Kuramashi:1993ka,Fukugita:1994ve,
Yamazaki:2004qb,Sasaki:2010zz,Aoki:2007rd,Prelovsek:2010kg,
Lang:2011mn,Feng:2010es,Meng:2009qt,He:2005ey,Meng:2008zzb}.

For some cases, we have to use the generic two-particle system
to extract the resonance parameters in the moving frame.
However, all of these aforementioned formulae in the moving frame
can apply only to two identical particle system.
For example, to examine the behavior of the $\kappa$ resonance,
it is highly desired for us to investigate $\pi K$ scattering
in the moving frame.
For a generic two-particle system in the moving frame,
we should change the  Rummukainen-Gottlieb's formulae,
which is devoted to the system of two identical particles.
To this purpose, we will strictly derive the equivalents of
the Rummukainen-Gottlieb's formulae
for a generic two-particle system in the moving frame
not only from theoretical aspects, but also from practical considerations.
This is very helpful for the lattice study
since it provides a feasible method
in the study of the $\kappa$ decay, vector kaon $K^*$ decay, and so on.

The alterations of the Rummukainen-Gottlieb's formulae
are mainly relevant with the different symmetries of
two-particle system in a torus~\footnote{
It is Naruhito Ishizuka who first help us to find the correct symmetry of
the two-particle system with unequal mass.
In there, we especially thank him.
Without his kind help, we can not finish this work smoothly.
}.
The representations of the rotational group $O(h)$
are decoupled into irreducible representations
of the $D_{4h}$ and $D_{2h}$ cubic groups for the system of
two identical particles with the non-zero total momentum
in a torus~\cite{Rummukainen:1995vs}.
In a generic two-particle system,
the symmetry is reduced.
For the case of ${\mathbf d}=(0,0,1)$,
the basic group is $C_{4v}$ instead of $D_{4h}$;
As for ${\mathbf d}=(0,1,1)$, the basic group becomes $C_{2v}$.
Hence, the final finite size formula for the generic two-particle system
in the moving frame is certainly new.

This paper is organized as follows. In Sec.~\ref{sec:torus},
we discuss the basic properties of the generic two-particle states in a torus.
In Sec.~\ref{sec:Thy_phaseshift}, we extend Rummukainen-Gottlieb's formulae
to the generic case  and derive the fundamental relationship for
the scattering phase in Eq.~(\ref{zetafunction_MF}),
and in Sec.~\ref{sec:thy_Symmetry}
we present the symmetry considerations.
Finally we give our brief conclusions in Sec.~\ref{sec:Conclusions}.

\section{Generic two-particle system on a cubic box}
\label{sec:torus}
Here we derive the formalisms required
for calculating the scattering phases in a cubic box,
which are enough for studying lattice simulation data.
However, in concrete lattice calculation,
we should address the lattice artifacts~\cite{Fu:2011xw}.
In this section we follow the Rummukainen and Gottlieb's
notations and conventions~\cite{Rummukainen:1995vs},
and extend them to the generic two-particle systems.

Without loss of generality, we consider two particles
with masses $m_1$ and $m_2$ for particle $1$ and particle $2$, respectively.
In this work we are specially interested in a system
having a non-zero total momentum,
namely, the moving frame~\cite{Rummukainen:1995vs}.
Using a moving frame with total momentum
${\mathbf P}=(2\pi/L){\mathbf d}$, ${\mathbf d}\in\mathbb{Z}^3$,
the energy eigenvalues for two-particle system in the non-interacting case
are given by~\cite{Rummukainen:1995vs}
$$
E_{MF} = \sqrt{m_1^2+p_1^2} + \sqrt{m_2^2+ p_2^2} ,
$$
where $p_1=|{\mathbf p}_1|$,
      $p_2=|{\mathbf p}_2|$,
and ${\mathbf p}_1$, ${\mathbf p}_2$ denote
the three-momenta of the particle $1$  and particle $2$, respectively,
which obey the periodic boundary condition,
$$
{\mathbf p}_i= \frac{2\pi}{L}{\mathbf n}_i ,
\quad {\mathbf n}_i\in \mathbb{Z}^3 ,
$$
and the relation
\begin{equation}
\label{eq:sum_p1p2_P}
\quad{\mathbf p}_1+{\mathbf p}_2={\mathbf P} .
\end{equation}

In the center-of-mass (CM) frame,
the total CM momentum disappears, namely,
$$
p^*=| {\mathbf p}^{\hspace{-0.05cm}*}|\,,\quad
{\mathbf p}^{\hspace{-0.05cm}*}={\mathbf p}^*_1=-{\mathbf p}^*_2 ,
$$
where ${\mathbf p}^{\hspace{-0.05cm}*}=(2\pi/L){\mathbf n}$, and
${\mathbf n}\in \mathbb{Z}^3$.
Here and hereafter we define the CM momenta with an asterisk $(^\ast)$.
The possible energy eigenstates of
the generic two-particle system are given by
$$
E_{CM} = \sqrt{m_1^2+ p^{*2}} + \sqrt{m_2^2 + p^{*2}} .
$$

The relativistic four-momentum squared is invariant,
and $E_{CM}$ is related to $E_{MF}$ in the moving frame
through the Lorentz transformation
$$
E_{CM}^2 = E_{MF}^2-{\mathbf P}^2 .
$$
In the moving frame, the center-of-mass is moving
with a velocity of ${\mathbf v}={\mathbf P}/E_{MF}$.
Using the standard Lorentz transformation
with a boost factor $\gamma=1/\sqrt{1-{\mathbf v}^2}$,
the $E_{CM}$ can be obtained through $E_{CM} = \gamma^{-1}E_{MF}$,
and momenta ${\bf p}_i$ and ${\bf p}^*$ are related by
the standard Lorentz transformation,
\begin{equation}
\label{eq:c_Lorentz_trans}
{\mathbf p}_1 =  \vec{\gamma}({\mathbf p}^* + {\mathbf v} E_1^*) \,, \quad
{\mathbf p}_2 = -\vec{\gamma}({\mathbf p}^* - {\mathbf v} E_2^*) \,,
\end{equation}
where $E_1^*$ and $E_2^*$ are energy eigenvalues of the particle $1$
and particle $2$ in the center-of-mass frame, respectively,
\begin{eqnarray}
\label{eq:Energy_C}
E_1^* &=& \frac{1}{2E_{CM}} \left( E_{CM}^2+m_1^2 -m_2^2\right) , \cr
E_2^* &=& \frac{1}{2E_{CM}} \left( E_{CM}^2+m_2^2 -m_1^2\right) ,
\end{eqnarray}
and the boost factor acts in the direction of ${\bf v}$,
here and hereafter we adopt the shorthand notation
\begin{equation}
\label{short_gamma}
\vec{\gamma}{\mathbf p} =
\gamma{\mathbf p}_{\parallel}+{\mathbf p}_{\perp}\,,\quad
\vec{\gamma}^{-1}{\mathbf p} =
\gamma^{-1}{\mathbf p}_{\parallel}+{\mathbf p}_{\perp} ,
\end{equation}
where ${\mathbf p}_{\parallel}$ and ${\mathbf p}_{\perp}$ are
the  ${\mathbf p}$ components which are parallel and perpendicular to
the CM velocity, respectively, namely,
\begin{equation}
{\mathbf p}_{\parallel}=
\frac{{\mathbf p}\cdot{\mathbf v}}{|\mathbf v|^2}{\mathbf v}\,,\quad
{\mathbf p}_{\perp}={\mathbf p}-{\mathbf p}_{\parallel}\,.
\end{equation}
Thus,  by inspecting Eqs.~(\ref{eq:sum_p1p2_P}), (\ref{eq:c_Lorentz_trans})
and (\ref{eq:Energy_C}), it can be seen that the ${\mathbf p}^*$ are quantized to the values
$$
{\mathbf p}^* =\frac{2\pi}{L}{\mathbf r},
\qquad {\mathbf r}\in P_{\mathbf d},
$$
where the set $P_{\mathbf d}$  is
\begin{equation}
\label{eq:set_Pd}
P_{\mathbf d} = \left\{ {\mathbf r} \left|  {\mathbf r} = \vec{\gamma}^{-1}
\left[{\mathbf n}+\frac{{\mathbf d}}{2} \cdot
\left(1+\frac{m_2^2-m_1^2}{E_{CM}^2}\right)
\right] \right. ,\; {\mathbf n}\in\mathbb{Z}^3 \right\} .
\end{equation}

In the interacting case, the $\overline{E}_{CM}$ is given by
\begin{equation}
\label{eq:continuum_dis_rel}
\overline{E}_{CM} = \sqrt{m_1^2 + k^{2}} + \sqrt{m_2^2 + k^{2}} ,
\quad k = \frac{2\pi}{L} q .
\end{equation}
where $q$ is no longer required to be an integer.
Solving this equation for scattering momentum $k$ we get
\begin{equation}
\label{eq:MF_k}
k = \frac{1}{2\overline{E}_{CM}}\sqrt{
[\overline{E}_{CM}^2 - (m_1 \hspace{-0.03cm}-\hspace{-0.03cm} m_2)^2]
[ \overline{E}_{CM}^2 -  (m_1 + m_2)^2] } .
\end{equation}
We can rewrite Eq.~(\ref{eq:MF_k}) to an elegant form as
\begin{equation}
\label{eq:MF_k_e_1}
k^2  = \frac{\overline{E}_{CM}^2}{4} +
\frac{ (m_1^2 - m_2^2)^2 }{4\overline{E}_{CM}^2} - \frac{m_1^2 + m_2^2}{2} .
\end{equation}

It is exactly this energy shift between the non-interacting situation
and interacting case, namely,
$\overline{E}_{CM} - E_{CM}$ (or equivalently $|{\mathbf n}|^2 - q^2$),
that we can evaluate two particle scattering phase.

As it is done in Ref.~\cite{Rummukainen:1995vs},
in the current study, we mainly investigate two moving frames.
One is ${\mathbf d}=(0,0,1)$,
where energy eigenstates transform under the tetragonal group $C_{4v}$,
only the irreducible representation $A_1$ is relevant for
two-particle scattering states in a cubic box with angular momentum $l=0$.
Another one is ${\mathbf d}=(0,1,1)$,
where energy eigenstates transform under the tetragonal group $C_{2v}$,
only the irreducible representation $A_1$ is relevant.
For the other cases, like ${\mathbf d}=(1,1,1)$, etc.,
we can easily work out from same way without difficulty.

Assuming that the phase shifts $\delta_l$ with
$l=1,2,3,\ldots$ are tiny in our concerned energy range,
the $s$-wave phase shift $\delta_0$ is connected to the scattering momentum $k$ by
\begin{equation}
\label{eq:Luscher_MF}
\tan\delta_0(k)=
\frac{\gamma\pi^{3/2}q}{\mathcal{Z}_{00}^{\mathbf d}(1;q^2)} ,
\end{equation}
where $k = (2\pi/L) q$, and the modified zeta function is defined by
\begin{equation}
\mathcal{Z}_{00}^{{\mathbf d}} (s; q^2) = \sum_{{\mathbf r}\in P_{\rm d}}
\frac{1} { ( |{\mathbf r}|^2 - q^2)^s } ,
\label{zetafunction_MF}
\end{equation}
and the set $P_{\mathbf d}$  is
\begin{equation}
\label{eq:set_Pd_Zeta}
P_{\mathbf d} = \left\{ {\mathbf r} \left|  {\mathbf r} =
\vec{\gamma}^{-1} \left[{\mathbf n}+\frac{{\mathbf d}}{2} \cdot
\left(1+\frac{m_2^2 \hspace{-0.05cm}-\hspace{-0.05cm} m_1^2}{E_{CM}^2}\right)
\right] , \right.  {\mathbf n}\in\mathbb{Z}^3 \right\} .
\end{equation}
For Eq.~(\ref{eq:Luscher_MF}), we note that the almost same result
has already existed in Eq.~(1) of Ref.~\cite{Davoudi:2011md},
where the formula was just presented without any explanation.
We can view our work as further confirming and strictly proving this formula.
The modified zeta function converges when $\mbox{Re}\,2s > l+3$,
and it can be analytically continued to whole complex plane.
The scattering momentum $k$ is defined from the invariant mass $\sqrt{s}$ as
$\sqrt{s} = \overline{E}_{CM} = \sqrt{ k^2 + m_1^2 } + \sqrt{ k^2 + m_2^2 }$.
The calculation method of $\mathcal{Z}_{00}^{{\mathbf d}} (1; q^2)$
is discussed in Appendix~\ref{app:zeta_lm} and in Ref.~\cite{Fu:2011xw}.
Using Eq.~(\ref{eq:Luscher_MF}), we can obtain the phase shift
from the energy spectrum using lattice simulations.
If we now set $m_1=m_2$, all the results in Ref.~\cite{Rummukainen:1995vs}
are elegantly recovered.

\section{Derivation of the phase shift formula}
\label{sec:Thy_phaseshift}
Here we deduce the basic  phase shift formula
in Eq.~(\ref{eq:Luscher_MF}) for the generic two-particle system with spin-$0$.
We utilize the Rummukainen and Gottlieb's formulae in Ref.~\cite{Rummukainen:1995vs},
and extend them to the generic two-particle system.
To make the derivation simple,
we are studying the system by the relativistic quantum mechanics.

Throughout this section, we employ the metric tensor
sign convention $g_{\mu\nu} = {\rm diag}(1,-1,-1,-1)$,
write the scalar productions in a compact form
$p^2 = p\cdot p  = p_\mu p^\mu$, etc,
and express the quantities in natural units with $\hbar = c = 1$.
Here and hereafter we follow the original notations
in Refs.~\cite{Rummukainen:1995vs}.

\subsection{Lorentz transformation of wave function}
Let us consider the generic system of two spin-$0$ particles
with mass $m_1$, and $m_2$, respectively, in a cubic box.
The two-particle system is described by the scalar wave
function $\psi(x_1,x_2)$, where $x_i = (x^0_i,{\mathbf x}_i), \ i=1,2$
are the four-dimensional Minkowski coordinates of two particles.
The wave function changes under the standard Lorentz transformations, namely,
\begin{equation}
\psi(x_1,x_2) = \psi^{\prime}(x_1^{\prime}, x_2^{\prime}) =
\psi^{\prime}(\Lambda x_1,\Lambda x_2),
\label{transform_psi}
\end{equation}
where $(x^{\prime})^\mu = {\Lambda^\mu}_\nu x^\nu$ defines
the standard Lorentz transformation of the four-vector $x$.

We can make the problem easier through the special properties of
the CM frame.
First let us study two non-interacting  particles,
and the wave functions of the system obey the Klein-Gordon (KG) equations
\begin{eqnarray}
\left( \hat p_{1\mu} \hat {p_1}^\mu - m_1^2 \right) \psi(x_1,x_2) &=& 0 , \cr
\left( \hat p_{2\mu} \hat {p_2}^\mu - m_2^2 \right) \psi(x_1,x_2) &=& 0 ,
\label{KleinGordon}
\end{eqnarray}
where $\hat p_{i\mu},\ i=1,2$ is the four-momentum operator.
It is well-known, the problem simplifies if we  separate the  variables
under the transformations
\begin{eqnarray}
X &=& \frac{m_1 x_1 + m_2 x_2}{m_1 + m_2}   \label{defX} , \\
x &=& x_1 - x_2                             \label{defx} ,
\end{eqnarray}
where $X$ is the position of the CM,
and $x$ is the relative coordinate of two particles.
Let us restrict ourselves to the solutions
which are the eigenstates of the CM  momentum operator.
Then Eq.~(\ref{KleinGordon}) can be changed into
\begin{eqnarray}
\hspace{-0.5cm}
\left[ \frac{m_1^2}{M^2} \hat P_\mu \hat P^\mu +
\hat p_\mu \hat p^\mu -  \frac{2m_1}{M} \hat p_\mu \cdot \hat P^\mu
 + m_1^2 \right] \psi(x,X) \hspace{-0.2cm} &=& \hspace{-0.2cm}0  , \label{xX_1} \\
\hspace{-0.5cm}
\left[ \frac{m_2^2}{M^2} \hat P_\mu \hat P^\mu +
\hat p_\mu \hat p^\mu + \frac{2m_2}{M} \hat p_\mu \cdot \hat P^\mu
 + m_2^2  \right] \psi(x,X) \hspace{-0.2cm}&=& \hspace{-0.2cm}0  , \label{xX_2}
\label{pdotP}
\end{eqnarray}
where
\begin{eqnarray}
\hat p &=& \frac{m_2 \hat p_1 - m_1 \hat p_2}{m_1 + m_2}\,, \label{def_p} \\
\hat P &=& \hat p_1 + \hat p_2 \,,\label{def_P} \\
M      &=& m_1 + m_2              \label{def_M} ,
\end{eqnarray}
$\hat p $ is relative four-momentum operator,
$\hat P$ is total four-momentum operator, and
$M$ is total mass of two particles.
Adding $1/m_1 \times$(\ref{xX_1}) to $1/m_2 \times$(\ref{xX_2}) and
subtracting (\ref{xX_1}) from  (\ref{xX_2}), respectively, yield
\begin{eqnarray}
\hspace{-0.8cm}\left[ \frac{ M^2} {m_1 m_2} \hat p_\mu \hat p^\mu - M^2
+ \hat P_\mu \hat P^\mu  \right] \psi(x,X)
\hspace{-0.1cm}&=&\hspace{-0.1cm} 0 \,,\label{pP_1}\\
\hspace{-0.8cm}\left[ \hat p_\mu \hat P^\mu -
\frac{m_1 - m_2}{2 M} \hat P_\mu \hat P^\mu -
\frac{m_1^2 - m_2^2}{2}  \right] \psi(x,X)
\hspace{-0.1cm}&=& \hspace{-0.1cm}0 \label{pP_2} \,.
\end{eqnarray}

It is well-known that, without external potentials,
the total momentum of the two-particle system is conserved,
thus we can restraint ourselves to the eigenfunctions of $P$, namely,
\begin{equation}
\psi(x,X) = e^{- i P_\mu X^\mu} \phi(x) ,
\label{ansatz_psi}
\end{equation}
where $P_\mu$ is a constant time-like vector,
and $P$ is denoted through $ P^2 = P_\mu P^\mu$.

In the present study, we are specially interested
in the CM frame, which is denoted as the frame
without the spatial components of the total momentum for
the system, namely, ${\mathbf P}^*=0$.
Thus, we can only take the positive kinetic energy
solutions $P^*_0 = E_{\rm CM}^* > m_1 + m_2$ into consideration.
So, Eqs.~(\ref{pP_1}) and (\ref{pP_2}) can be rewritten as,
\begin{eqnarray}
\hspace{-0.6cm}
\left( \hat p_\mu^* \hat p^{*\mu}  + \frac{E_{\rm CM}^2 m_1 m_2}{(m_1+m_2)^2}
- m_1 m_2 \right) \phi_{\rm CM}(x^*) &=& 0 , \label{qQ_1}\\
\hspace{-0.6cm}
\left( \hat p^*_0  - \frac{E_{\rm CM}}{2} \frac{m_1 - m_2}{m_1+m_2}
-\frac{m_1^2 - m_2^2}{ 2E_{\rm CM}}  \right) \phi_{\rm CM}(x^*)  &=&  0 \label{qQ_2} .
\end{eqnarray}
Eq.~(\ref{qQ_2}) indicates $\hat p^*_0 \phi_{\rm CM}(x^*) \ne 0$ for $m_1\ne m_2$.
By inspecting Eq.~(\ref{qQ_1}) and Eq.~(\ref{qQ_2}),
we can reasonably assume that the wave function $\phi_{\rm CM}(x^*)$
can be expressed as
\begin{equation}
\phi_{\rm CM}(x^*) \equiv e^{ i\beta x^{\hspace{-0.05cm}*0} }
\phi_{\rm CM}( {\mathbf x}^*) ,
\label{eq_seq_CM_wF}
\end{equation}
where $x^{\hspace{-0.05cm}*0}=x_1^{*0} - x_2^{*0}$ is
the relative temporal separation of two particles,
and $\beta$ is a constant, namely,
\begin{equation}
\beta = \frac{E_{\rm CM}}{2} \frac{m_2 - m_1}{m_1+m_2}+
\frac{m_2^2 - m_1^2}{ 2E_{\rm CM}}  .
\label{beta_factor}
\end{equation}
It is obvious that when $m_1=m_2$, $\beta \to 0$.
So, in the CM frame the wave function
depends explicitly on the time variable
$t^* \equiv X^{\hspace{-0.05cm}*0} =
( m_1 x_1^{*0} + m_2 x_2^{*0})/(m_1 + m_2)$,
the relative spatial separation
${\mathbf x}^*={\mathbf x}^*_1 -{\mathbf x}^*_2$,
and the relative temporal separation of the particles
$x^{\hspace{-0.05cm}*0}$, namely,
\begin{equation}
\psi_{\rm CM}(x^*, t^{*}) = e^{-i E_{\rm CM} t^{*}}
e^{ i\beta x^{\hspace{-0.05cm}*0} } \phi_{\rm CM}({\mathbf x}^*) ,
\label{psi_cm}
\end{equation}
where the constant $\beta$ is denoted in  Eq.~(\ref{beta_factor}).

Now let us discuss the case in the moving frame.
The transformation from the MF to CM frame
can be expressed as $r^{*\mu} = {\Lambda^\mu}_\nu r^\nu$,
where $r$ is any position four-vector and
quantities without $*$ stand for these in the moving frame.
Using the notation in~(\ref{short_gamma}), we have
\begin{equation}
r^{*0}         = \gamma (r^0 + {\mathbf v}\cdot{\mathbf r}) ,  \quad
{\mathbf r}^*  = \vec\gamma\, ({\mathbf r} + {\mathbf v} r^0) ,
\end{equation}
where $\gamma$ is a boost factor, and ${\mathbf v} = {\mathbf P}/P_0$ is
the three-velocity of the CM in the moving frame.
We can rewrite ${\mathbf v}$ to a form for later use as
\begin{equation}
{\mathbf v} = \frac{2\pi}{L E_{\rm MF}}{\mathbf d} =
\frac{2\pi}{\gamma L E_{\rm CM}}{\mathbf d} .
\label{eleg_V}
\end{equation}
Considering  the identity $P_\mu X^\mu = P^*_\mu X^{*\mu}$, the Lorentz transformation in Eq.~(\ref{transform_psi}), and Eq.~(\ref{ansatz_psi}),
the wave function in the moving frame can be expressed as
$$
\psi_{\rm MF}(x,X) = e^{-iP_\mu X^\mu} \phi_{\rm MF}(x) ,
$$
where
\begin{equation}
\phi_{\rm MF}(x) \equiv \phi_{\rm MF}(x^0,{\mathbf x}) =
\phi_{\rm CM}\left(\gamma (x^0 + {\mathbf v}\cdot{\mathbf x}),
              \vec\gamma\, ({\mathbf x} + {\mathbf v} x^0)
\right) .
\label{phi_l_func}
\end{equation}
Thus, the wave function $\phi_{\rm MF}$ depends on
time separation $x^0 = x_1^0 - x_2^0$ explicitly.
However, in the moving frame we only consider
two particles with the same time coordinate, namely, $x^0 = 0$.
It corresponds to the tilted plane $(x^{\hspace{-0.05cm}*0},{\mathbf x}^*) =
(\gamma {\mathbf v}\cdot{\mathbf x},\vec\gamma\,{\mathbf x})$
for the CM frame,
since the wave function $\phi_{\rm CM}$ is dependent of the relative temporal
separation $x^{\hspace{-0.05cm}*0}$,
we can clearly observe the effect of the tilt to the wave function,
and Eq.~(\ref{phi_l_func}) take as
\begin{equation}
\label{lfun_cm_rel}
\phi_{\rm MF}(0,{\mathbf x}) =
\phi_{\rm CM}(\gamma {\mathbf v}\cdot{\mathbf x}, \vec\gamma\,{\mathbf x}) .
\end{equation}
Using Eq.~(\ref{eq_seq_CM_wF}) and Eq.~(\ref{eleg_V}),
we can rewrite Eq.~(\ref{lfun_cm_rel}) as
\begin{equation}
\label{lfun_cm_rel_S}
\phi_{\rm MF}(0,{\mathbf x}) =
e^{i\beta^\prime \pi {\mathbf d}\cdot{\mathbf x}/L}
\phi_{\rm CM}(\vec\gamma\,{\mathbf x}) .
\end{equation}
where $\beta^\prime$ is a constant, namely,
\begin{equation}
\label{beta_factor1}
\beta^\prime = \frac{m_2 - m_1}{m_1+m_2}+
\frac{m_2^2 - m_1^2}{ E_{\rm CM}^2 }  .
\end{equation}
Eq.~(\ref{lfun_cm_rel_S}) has a simple physical meaning:
the CM system watches the torus in the moving frame
expanded by a boost factor $\gamma$ in the direction of ${\bf P}$,
and the length scales in perpendicular directions are hold.
At last, Eq.~(\ref{lfun_cm_rel_S}) connects the MF wave function,
\begin{equation}
\label{psi_lf}
\psi_{\rm MF}(0,{\mathbf x},t,{\mathbf X}) =
e^{-iE_{\rm MF} t + i{\mathbf P} \cdot{\mathbf X}} \phi_{\rm MF}(0,{\mathbf x}) ,
\end{equation}
to the CM frame wave function Eq.~(\ref{psi_cm}).
By inspecting Eqs.~(\ref{qQ_1}), (\ref{qQ_2}) and (\ref{psi_cm}),
we, at last, achieve the wave function $\phi_{\rm CM}$
meeting the Helmholtz equation (HE)
\begin{equation}
\label{phi_cm_eq}
(\nabla^2_{{\mathbf x}^{\hspace{-0.03cm}*}} + k^{\hspace{-0.05cm}*2})
\phi_{\rm CM}({\mathbf x}^{\hspace{-0.05cm}*}) = 0 ,
\end{equation}
where
\begin{equation}
\label{eq:MF_p_e_1}
k^{\hspace{-0.05cm}*2}  =
\frac{E_{CM}^2}{4} + \frac{(m_1^2 - m_2^2)^2}{4 E_{CM}^2}
- \frac{m_1^2 + m_2^2}{2} .
\end{equation}
This result is consistent with the solution in Ref.~\cite{Tam:1975ag}.

The Eqs.~(\ref{lfun_cm_rel_S}) and (\ref{phi_cm_eq})
are very important
when we study the wave functions of our system.
Thus Eq.~(\ref{phi_cm_eq}) is a very important result,
which represents one of the main results of the present work.
In the following studies, we take away the superscript $^*$
from the quantities in the CM frame.
We can easily check that if we take $m_1=m_2$,
all the corresponding results in Ref.~\cite{Rummukainen:1995vs} are restored.

\subsection{Modified singular ${\mathbf d}$-periodic solutions of the Helmholtz equation}
\label{sec:singular}
In our concrete problem,
for the potential $V_\mu({\mathbf x})$ with
a limit range~\cite{Luscher:1991cf}, namely,
\begin{equation}
\label{eq:potential}
V_\mu({\mathbf x}) = 0 \quad {\rm for} \,\, |{\mathbf x}| > R ,
\end{equation}
we suppose that the KG equation~(\ref{KleinGordon})
in the CM frame still has a square integral solution.
In the CM frame the interaction of the system
is spherically symmetric.
The wave function of the system is usually given in spherical harmonics
\begin{equation}
\label{eq:spherical}
\phi_{\rm CM}({\mathbf x}) = \sum_{l=0}^{\infty} \sum_{m=-l}^{l}
Y_{lm}(\theta,\varphi) \phi_{lm}(x) .
\end{equation}

For $x > R$, $\phi_{\rm CM}$ is a solution of HE,
and the radial functions $\phi_{lm}$ meet the differential equation
\begin{equation}
\label{eq:radial}
\left[\frac{\mbox{d}^2}{\mbox{d} x^2} +
\frac{2}{x} \frac{\mbox{d}}{\mbox{d} x}
- \frac{l(l+1)}{x^2} + k^{2} \right] \phi_{lm}(x) = 0 ,
\end{equation}
where
\begin{equation}
\label{eq:MF_pp_e_1}
k^{2}  =  \frac{E_{CM}^2}{4} + \frac{ (m_1^2 - m_2^2)^2}{4E_{CM}^2}
- \frac{m_1^2 + m_2^2}{2} .
\end{equation}
With the linear combinations of the spherical Bessel functions
the solutions of Eq.~(\ref{eq:radial}) can be expressed as
\begin{equation}
\label{eq:bessel}
\phi_{lm}(x) = c_{lm} \left[ a_l(k) j_l(k x) + b_l(k) n_l(k x) \right].
\end{equation}

Although we do not know the radial equation in the region $x < R$,
by comparing the wave functions denoted in Eqs.~(\ref{eq:spherical},\ref{eq:bessel}),
we can get the relation
between the scattering phase and the coefficients
$a_l$ and $b_l$~\cite{Luscher:1991cf}, namely,
$$
e^{i 2\delta_l(k)} = \frac{a_l(k) + ib_l(k)}{a_l(k) - ib_l(k)} .
$$
Because the radial equation $a_l$ and $b_l$ can be taken to be
the real number for $k > 0$, $\delta_l(k)$ is real.
Thus, for a given $l$-sector, the phase shift $\delta_l(k)$
can be expressed by  energy $E_{\rm MF}$ through
$$
k^{2}  =  \frac{E_{\rm MF}^2 - {\mathbf P}^2}{4} +
\frac{1}{4}\frac{(m_1^2 - m_2^2)^2}{E_{\rm MF}^2 - {\mathbf P}^2}
- \frac{m_1^2 + m_2^2}{2} .
$$

In the moving frame, we now investigate two generic particles
confined in a torus of size $L^3$ with periodic boundary conditions(PBC).
The temporal direction of the torus is chosen to be infinite.
The wave functions $\psi_{\rm MF}$ in MF should be
periodic with regard to the position of each particle, namely,
\begin{equation}
\label{lfunc_period}
\psi_{\rm MF}({\mathbf x}_1,{\mathbf x}_2) =
\psi_{\rm MF}({\mathbf x}_1 + {\mathbf l}L,{\mathbf x}_2 + {\mathbf m}L) ,
\quad {\mathbf l},{\mathbf m} \in {\mathbb Z}^3 .
\end{equation}
The wave function $\psi_{\rm MF}$ is provided by~(\ref{psi_lf}), namely
\begin{equation}
\label{lf_ansatz}
\psi_{\rm MF}({\mathbf x}_1,{\mathbf x}_2) =
\exp \left( {i\frac{ {\mathbf P}\cdot(m_1 {\mathbf x}_1+ m_2{\mathbf x}_2)}{m_1+m_2} } \right)
\phi_{\rm MF}({\mathbf x}_1 - {\mathbf x}_2).
\end{equation}
Combining Eq.~(\ref{lfunc_period}) and Eq.~(\ref{lf_ansatz}) gives
\begin{eqnarray}
{\mathbf P} &=& {\displaystyle \frac{2\pi}{L}}\,{\mathbf d} ,  \\
\phi_{\rm MF}({\mathbf x}) &=&
{ e^{i\pi\frac{ 2m_1}{m_1+m_2} {\mathbf d}\cdot{\mathbf n}} }
\phi_{\rm MF}({\mathbf x}+{\mathbf n}L) ,
\label{lf_quant}
\end{eqnarray}
where ${\mathbf n}= {\mathbf l} -{\mathbf m}$,
${\mathbf d},{\mathbf n} \in {\mathbb Z}^3$  and
${\mathbf P}$ is the total momentum~\footnote{
Eq.~(\ref{lf_quant}) can also be written as
$$\phi_{\rm MF}({\mathbf x}) =
{ e^{-i\pi\frac{ 2m_2}{m_1+m_2} {\mathbf d}\cdot{\mathbf n}} }
\phi_{\rm MF}({\mathbf x}+{\mathbf n}L) \,.
$$
Following the almost same procedures and addressing the corresponding formulae,
we can arrive at the same final numerical results.
For our case, ${\mathbf d}\cdot{\mathbf n}$ is an integer.
}.
The rule (\ref{lf_quant}) separates the wave functions
into the different sectors, which can be categorized
by the three-vector ${\mathbf d}$.
In the current study, we naturally concentrate on
sectors ${\mathbf d}=(0,0,1)$ and ${\mathbf d}=(0,1,1)$.

We are now on the position to employ Eq.~(\ref{lfun_cm_rel_S}) to
get the corresponding periodic rule
for the CM wave function.
For a chosen vector ${\mathbf d}$, we have
\begin{equation}
\phi_{\rm CM}({\mathbf x}) =
e^{i\pi {\mathbf d}\cdot{\mathbf n}\left( 1+\frac{m_2^2 - m_1^2}{ E_{\rm CM}^2 }\right) }
\phi_{\rm CM}({\mathbf x}+\vec\gamma\,{\mathbf n}L) ,
\quad {\mathbf n}\in {\mathbb Z}^3.
\label{cm_period}
\end{equation}
For simpleness, we refer to the functions
obeying the periodicity rule~(\ref{cm_period}) as
modified ${\mathbf d}$-periodic rule.
As shown later, the modified ${\mathbf d}$-periodic rule~(\ref{cm_period})
is a milestone in this work.

Now we deduce the general solutions of
the HE satisfying the modified ${\bf d}$-periodic rule~(\ref{cm_period}).
Except the modified ${\mathbf d}$-periodicity,
our derivation follows the works in section $4.4$
of Ref.~\cite{Rummukainen:1995vs}.
In this work, we will refer to a function $\phi$ as a singular
modified ${\mathbf d}$-periodic solution of the HE,
if it is a smooth function denoted for
all ${\mathbf x}\neq \vec\gamma\,{\mathbf n}L$, ${\mathbf n}\in{\mathbb Z}^3$,
and it meets the HE, namely,
$$
(\nabla^2 + k^{2}) \phi({\mathbf x})=0 ,
$$
for $k>0$, and meets the modified ${\mathbf d}$-periodic rule,i.e.,
\begin{equation}
\label{m_d_period}
\phi({\mathbf x}) = e^{i\alpha\pi{\mathbf d}\cdot{\mathbf n}}
\phi({\mathbf x} + \vec\gamma\,{\mathbf n} L) ,
\quad  {\mathbf n}\in{\mathbb Z}^3 ,
\end{equation}
here and hereafter, for compactness, we defined a factor $\alpha$, namely,
\begin{equation}
\alpha  \equiv 1+\frac{m_2^2-m_1^2}{E_{CM}^2} .
\label{alpha_factor}
\end{equation}
When $m_1 = m_2$, $\alpha=1$, this is the case
Rummukainen and Gottlieb studied in Ref.~\cite{Rummukainen:1995vs}.
Moreover, the wave function is required to be bounded by a power of
$1/|{\mathbf x}|$ around the origin, namely
$$
\lim_{{\mathbf x}\rightarrow 0} |{\mathbf x}^{\Lambda+1} \phi({\mathbf x}) |
< \infty
$$
for the positive integer $\Lambda$, the degree of $\phi$.
For our aim, it is enough to study the regular values of $k$, namely
$$
k \neq \frac{2\pi}{L}\left|\vec\gamma^{-1}
\left({\mathbf n} + \frac{\alpha }{2} {\mathbf d} \right)  \right| ,
\quad {\mathbf n}\in{\mathbb Z}^3.
$$

We denote the Green function
\begin{equation}
\label{eq:green}
G^{\mathbf d}({\mathbf x};k) =
\gamma^{-1} L^{-3} \sum_{{\mathbf p}\in\Gamma}
\frac{e^{i{\mathbf p}\cdot{\mathbf x}}}{{\mathbf p}^2 - k^{2}} ,
\end{equation}
where
$$
\Gamma = \left\{{\mathbf p}\in{\mathbb R}^3 \,\Big|\,
{\mathbf p} = \frac{2\pi}{L} \vec\gamma^{-1}
\left({\mathbf n} + \frac{\alpha  }{2} {\mathbf d} \right) ,
\quad {\mathbf n}\in{\mathbb Z}^3 \right\} .
$$
Eq.~(\ref{eq:green}) is well-defined due to the non-singular $k$.
If we select
${\mathbf k} =
(2\pi/L)\vec\gamma^{-1}\left({\mathbf m} +
\frac{\alpha}{2}{\mathbf d}\right), \, {\mathbf m}\in{\mathbb Z}^3$,
then
$$
{\mathbf k}\cdot({\mathbf x} + \vec\gamma {\mathbf n} L) =
{\mathbf k}\cdot{\mathbf x} +
\alpha \pi{\mathbf d}\cdot{\mathbf n} +
2\pi{\mathbf m}\cdot{\mathbf n} ,
$$
where $ {\mathbf n}\in{\mathbb Z}^3$,
and the Green function $G^{\mathbf d}({\mathbf x};k)$ meets clearly
the modified ${\mathbf d}$-periodic rule, as we expected.
Furthermore, it obeys
$$
(\nabla^2 + k^{2})G^{\mathbf d}({\mathbf x};k) =
-\sum_{{\mathbf n}\in{\mathbb Z}^3}
e^{i\alpha \pi{\mathbf d}\cdot{\mathbf n}}
\delta({\mathbf x}+ \vec\gamma\,{\mathbf n} L) .
$$
We can easily verify that the function $G^{\mathbf d}({\mathbf x};k)$ is a
singular modified ${\mathbf d}$-periodic solution
of Helmholtz equation with degree $1$.
Other singular periodic solutions can be easily achieved by
differentiating $G^{\mathbf d}$ with ${\mathbf x}$.
And
\begin{equation}
G^{\mathbf d}_{lm}({\mathbf x};k) =
{\mathcal Y}_{lm}(\nabla)G^{\mathbf d}({\mathbf x};k) ,
\end{equation}
where the harmonic polynomials
${\mathcal Y}_{lm}({\mathbf x}) = x^lY_{lm}(\theta,\varphi)$.
We can easily verify that the functions $G^{\mathbf d}_{lm}$ are
singular modified ${\mathbf d}$-periodic solutions of the HE,
and they make up a complete set of solutions
so that any singular modified ${\mathbf d}$-periodic solution
of degree $\Lambda$ is just a linear combination of
$G^{\mathbf d}_{lm}({\mathbf x};p)$ for $l\le\Lambda$~\cite{Luscher:1991cf}.
In the region $0<x<L/2$ the functions $G^{\mathbf d}_{lm}$
can be expanded in usual spherical harmonics, namely
\begin{eqnarray}
\label{G_Spherical}
\hspace{-0.5cm}G^{\mathbf d}_{lm}({\mathbf x};k) &=&
\frac{(-1)^l k^{l+1}}{4\pi} \left[ n_l(kx)Y_{lm}(\theta,\varphi) \right. + \cr
&&  \hspace{-0.7cm}
\left.\sum_{l^{\prime}=0}^{\infty} \sum_{m^{\prime}=-l}^{l}
{\mathcal M}^{{\mathbf d}}_{lm,l^{\prime}m^{\prime}}(k) \,
j_{l^{\prime}}(kx) Y_{l^{\prime}m^{\prime}}(\theta,\varphi)
\right] .
\end{eqnarray}
The regular part has the coefficients
${\mathcal M}^{{\mathbf d}}_{lm,l^{\prime}m^{\prime}}(k)$.
In the concrete calculation, only a few of them is needed,
 and we also provide its general expression, namely,
\begin{equation}
\label{eq:M_expression}
{\mathcal  M}^{{\mathbf d}}_{lm,l^{\prime}m^{\prime}}(k) =
\frac{(-1)^l}{\gamma\pi^{3/2}}
\sum_{j=|l-l^{\prime}|}^{l+l^{\prime}}
\sum_{s=-j}^j \frac{i^j}{q^{j+1}}
{\mathcal Z}_{js}^{{\mathbf d}} (1;q^2)
C_{lm,js,l^{\prime}m^{\prime}} ,
\end{equation}
where $ q = kL/(2\pi)$.
Using Wigner $3j$-symbols~\cite{Rotenberg}
the tensor $C_{lm,js,l^{\prime}m^{\prime}}$ can be expressed as
\begin{eqnarray}
\label{eq:C_Expression}
C_{lm,js,l^{\prime}m^{\prime}} &=& (-1)^{m^{\prime}} i^{l-j+l^{\prime}}
\sqrt{(2l+1)(2j+1)(2l^{\prime}+1)} \cr
&\times&   \left(
\begin{array}{rrr}
l & j & l^{\prime} \\
m & s & -m'
\end{array} \right)
\left(
\begin{array}{rrr}
l & j & l^{\prime} \\
0 & 0 & 0
\end{array}
\right)  .
\end{eqnarray}
The modified zeta function is formally denoted by
\begin{equation}
{\mathcal Z}_{lm}^{{\mathbf d}} (s; q^2) = \sum_{{\mathbf r}\in P_{\mathbf d}}
\frac{ {\mathcal Y}_{lm}({\mathbf r}) }
     { ({\mathbf r}^2 - q^2)^s } ,
\label{zeta_Func}
\end{equation}
where
$$
P_{\mathbf d} =  \left\{
{\mathbf r}\in{\bf R}^3\,\left|\,{\mathbf r} = \vec\gamma^{-1}
\left({\mathbf n} + \frac{\alpha}{2} {\mathbf d}  \right),
\quad {\mathbf n}\in {\mathbb Z}^3
\right\} .
\right.
$$

In Table~\ref{table:Mtable} we summarized the expressions of
${\mathcal M}_{lm,l^{\prime}m^{\prime}}^{\mathbf d}$ for $l,l^{\prime}\le 3$.
For compactness, we denoted
\begin{equation}
w_{lm} = \frac{1}{\pi^{3/2}\sqrt{2l+1}}
\gamma^{-1}q^{-l-1} {\mathcal Z}_{lm}^{\mathbf d}(1;q^2) .
\label{eq:wlm}
\end{equation}
The Wigner $3j$-symbol values
can be obtained in Ref.~\cite{Rotenberg}.
\begin{table*}[htb!]
\begin{ruledtabular}
\begin{tabular}{rrrrr|rrrrrrrrr}
$l$ & $m$ & $l^{\prime}$ & $m^{\prime}$ & &
	\multicolumn{9}{c}{$M_{lm,l^{\prime}m^{\prime}}^{\mathbf d}$} \\
\hline
 0& 0& 0& 0 && $w_{00}$&&&&&&& \\
\hline
 1 & 0 & 0 & 0 &&        & $iw_{10}$&& & & & & &  \\
 1 & 0 & 1 & 0 &&$w_{00}$&&$           + 2 w_{20}$& && & & & \\
 1 & 1 & 1 & 1 &&$w_{00}$&&$             - w_{20}$& && & & & \\
\hline
2 & 0 & 0 & 0 &&$      $&&$     -\sqrt{5} w_{20}$& & && &  \\
2 & 0 & 1 & 0 &&$       $&$i\sqrt{\frac{4}{5}}w_{10}$& & $ i\sqrt{\frac{27}{35}}w_{30}$& && & &\\
2 & 1 & 1 & 1 &&$       $&$i\sqrt{\frac{1}{5}}w_{10}$& & $ i\sqrt{\frac{18}{35}}w_{30}$& && & &\\
2 & 0 & 2 & 0 &&$w_{00}$&&$+ \frac{10}{7}w_{20}$& &$+\frac{18}{7} w_{40}$& & & &\\
2 & 1 & 2 & 1 &&$w_{00}$&&$+ \frac{5}{7}w_{20}$ & &$-\frac{12}{7} w_{40}$& & & &\\
2 & 2 & 2&$-2$&& && & &$ \frac{3}{7}\sqrt{70}w_{44}$& &&& \\
2 & 2 & 2 & 2 &&$w_{00}$&&$  - \frac{10}{7} w_{20}$&&$+ \frac{3}{7} w_{40}$&&&& \\
\hline
 3 & 0 & 0 & 0 &&&&& $-iw_{30}$&  & & & & \\
 3 & 0 & 1 & 0 &&$      $&&$-\frac{3}{7}\sqrt{21}w_{20}$&&$-\frac{4}{7}\sqrt{21}w_{40}$&& & & \\
 3 & 1 & 1 & 1 &&$      $&&$-\frac{3}{7}\sqrt{14}w_{20}$&&$+\frac{3}{7}\sqrt{14}w_{40}$&& & &\\
 3 & 3 & 1&$-1$&&$      $&&&&&$2\sqrt3 w_{44}$& &  \\
 3 & 0 & 2 & 0 &&&$ -i3\sqrt{\frac{3}{35}}w_{10}$ && $ -i\frac{4}{3}\sqrt{\frac{1}{5}}w_{30}$  & & &
 $ -i\frac{10}{9}\sqrt{\frac{1}{111}}w_{50}$  &   \\
 3 & 1 & 2 & 1 &&&$ -i2\sqrt{\frac{6}{35}}w_{10}$ && $ -i\sqrt{\frac{2}{105}}w_{30}$   & & &
 $ -i\frac{5}{9}\sqrt{\frac{2}{111}}w_{50}$  &\\
 3 & 2 & 2 & 2 &&&$ i\sqrt{\frac{3}{7}}w_{10}$ && $ i\frac{2}{3} w_{30}$   & & &
 $ i\frac{1}{3}\sqrt{\frac{5}{11}}w_{50}$  & \\
 3 & 0 & 3 & 0 &&$w_{00}$&&$       + \frac{4}{3} w_{20}$&&$+ \frac{18}{11} w_{40}$&&&
 $+\frac{100}{33} w_{60}$&  \\
 3 & 1 & 3 & 1 &&$w_{00}$&&$+w_{20}$&&$+  \frac{3}{11} w_{40}$&&&$- \frac{25}{11} w_{60}$&  \\
 3 & 2 & 3&$-2$&&$      $&&&&&$ \frac{3}{11}\sqrt{70} w_{44}$&&&$+\frac{10}{11}\sqrt{14}w_{64}$  \\
 3 & 2 & 3 & 2 &&$w_{00}$&&&&$-\frac{21}{11} w_{40}$&&&$+\frac{10}{11} w_{60}$&  \\
 3 & 3 & 3&$-1$&&$      $&&&&&$ \frac{3}{11}\sqrt{42} w_{44}$&&&$-\frac{5}{33}\sqrt{210}w_{64}$  \\
 3 & 3 & 3 & 3 &&$w_{00}$&&$       - \frac{5}{3} w_{20}$&&$+\frac{9}{11} w_{40}$&&&$-\frac{5}{33} w_{60}$& \\
\end{tabular}
\end{ruledtabular}
\caption[1]{\label{table:Mtable}
The matrix elements ${\mathcal M}_{lm,l^{\prime}m^{\prime}}^{\mathbf d}$
for ${\mathbf d}=(0,0,1), m_1\ne m_2$ and  $l,l^{\prime} \le 3$.
}
\end{table*}
We can easily verify that, if we set $m_1=m_2$, all of the above
definitions and formulae nicely reduce to the
those obtained in Ref.~\cite{Rummukainen:1995vs}, as we expected.
Of course, if we further select ${\mathbf d}=0$,
the moving frame and the CM frame coincide,
$\gamma\rightarrow 1$ and $P_{\mathbf d}\rightarrow{\mathbb Z}^3$,
and they further neatly reduce to that in Ref.~\cite{Luscher:1991cf}.
The Table~\ref{table:Mtable} can be compared with the Table $3$ in
Ref.~\cite{Rummukainen:1995vs},
which summaries the matrix elements for $m_1 = m_2$.
We should stress that the functions $w_{10}$, $w_{30}$
and $w_{50}$ appear in Table~\ref{table:Mtable}.
If we set $m_1=m_2$, then $w_{10}\rightarrow 0$, $w_{30}\rightarrow0$, and
$w_{50}\rightarrow 0$,
and Rummukainen-Gottlieb's results is elegantly restored.

So far, we can arrive at
\begin{eqnarray}
&&\hspace{-1.0cm}\sum_{l=0}^\Lambda \sum_{m=-l}^{l} v_{lm} G_{lm}^{{\mathbf d}} ({\mathbf x}, k^2) = \nonumber \\
&&\hspace{-1.0cm}\sum_{l=0}^\Lambda \sum_{m=-l}^{l} c_{lm}[a_l(k)j_l(kx) +
b_l(k)n_l(kx)]\,Y_{lm}(\theta,\varphi)
\end{eqnarray}
for the constants $c_{lm}$ and $v_{lm}$. And we obtain
\begin{equation}
\label{c-eq}
c_{lm}a_l(k) =  \sum_{l^{\prime}=0}^\Lambda
\sum_{m^{\prime}=-l^{\prime}}^{l^{\prime}}
c_{l^{\prime}m^{\prime}} b_{l^{\prime}}(k)
{\mathcal M}_{l^{\prime}m^{\prime},lm}^{{\mathbf d}}(k).
\end{equation}
${\mathcal M}_{l^{\prime}m^{\prime},lm}$ can be viewed
as the matrix element of the operator $M$.
We can express Eq.~(\ref{c-eq}) as a matrix equation,
$
C(A-BM) = 0 ,
$
where matrix $A_{(lm),(l^{\prime}m^{\prime})} =
a_l(p) \delta_{l,l^{\prime}}\delta_{m,m^{\prime}}$ (similar for $B$).
We can denote the phase shift matrix~\cite{Luscher:1991cf,Rummukainen:1995vs},
$$
e^{2i\delta} = \frac{A+iB}{A-iB} .
$$
The determinant condition arrives at~\cite{Rummukainen:1995vs}
\begin{equation}
\label{determinant}
\det \left[e^{2i\delta} (M-i) - (M+i)\right] = 0.
\end{equation}

\section{Symmetry discussions}
\label{sec:thy_Symmetry}
When the moving and CM frames coincide,
the two-particle system has a cubic symmetry and
the wave functions transforms
under the representations of the cubic group $O_h$.
On the other hand, according to Eq.~(\ref{lfun_cm_rel_S}),
if two frames are not equivalent,
the Lorentz translation boost from the moving frame
to the CM frame and only some subgroups of $O_h$ group survive~\cite{Rummukainen:1995vs}.

In this work, we are mainly interested in a boost along
one of the coordinate axes, namely ${\mathbf d}=(0,0,1)$.
The geometry of the torus transforms as $(1,1,1)\rightarrow (1,1,\gamma)$,
and the corresponding symmetry group is tetragonal point group $C_{4v}$,
which has $8$ elements: $4$ rotations
through an angle $(n\pi/2)$, where $n=0,1,2,3$, around the $x_3$-axis;
and all four of the above multiplied by the reflection with respect to the $(1,3)$-plane.
The relevant point groups and the boost vectors are classified
in Table~\ref{table:symmetry}.

\begin{table}[thb]
\begin{ruledtabular}
\begin{tabular}{cclc}
 ${\mathbf d}$ &  point group & classification & $N_{\rm elements}$ \\ \hline
 $(0,0,0)$     & $O_h$        & cubic          & 48  \\
 $(0,0,a)$     & $C_{4v}$     & tetragonal     & 8  \\
 $(0,a,a)$     & $C_{2v}$     & orthorhombic   & 4  \\
\end{tabular}
\end{ruledtabular}
\caption[1]{\label{table:symmetry}
The categorizations of the boosts on a cubic box and
its reduction of the cubic group.
The first column shows the boost direction and $a$ is a non-zero real number.
Here we adopted the Schonflies notation~\cite{Weissbluth}.
}
\end{table}

In this work we mainly discuss the cubic
and tetragonal symmetry groups $O_h$, $C_{4v}$, and $C_{2v}$.
The tetragonal group $C_{4v}$ has
four $1$-dimensional representations $A_1$, $A_2$, $B_1$, $B_2$,
and one $2$-dimensional representation $E$~\cite{Weissbluth}.
The representations of the rotational group are
reduced into irreducible representations of $C_{4v}$ as:
\begin{eqnarray}
\Gamma^{(0)} &=& A_1                                 \,, \cr
\label{eq:decomp_c4v}
\Gamma^{(1)} &=& A_1 \oplus E                        \,,\\
\Gamma^{(2)} &=& A_1 \oplus B_1  \oplus B_2 \oplus E \,. \nonumber
\end{eqnarray}

The representations can be obtained through
the character tables~\cite{Weissbluth} or
by enumerating harmonic polynomials of degree $l$
which transform under the representations of $C_{4v}$.
The basis polynomials for the corresponding representations
are summarized in Table~\ref{table:basis:c4v} for $l\le 2$.
\begin{table}[h]
\begin{ruledtabular}
\begin{tabular}{c|llll}
\mbox{representation} &  $l=0$ &  $l=1$  &  $l=2$   & {\mbox{indices}} \\
\hline
$A_1$ &$1$& $x_3$ & $x_3^2 - \frac{1}{3}x^2$ &         \\
$A_2$ &   &       &                          &         \\
$B_1$ &   &       & $x_1^2 - x_2^2$          &         \\
$B_2$ &   &       & $x_1x_2$                 &         \\
$E$   &   & $x_i$ & $x_i x_3$                & $i=1,2$ \\
\end{tabular}
\end{ruledtabular}
\caption{\label{table:basis:c4v}
The basis polynomials for the irreducible representations of $C_{4v}$.
}
\end{table}

The tetragonal group $C_{2v}$ has four $1$-dimensional representations
$A_1$, $A_2$, $B_1$, $B_2$~\cite{Weissbluth}.
The representations of the rotational group are further
reduced into irreducible representations of $C_{2v}$ as:
\begin{eqnarray}
\Gamma^{(0)} &=& A_1                                   , \cr
\label{eq:decomp_c2v}
\Gamma^{(1)} &=& A_1 \oplus B_1 \oplus B_2             , \\
\Gamma^{(2)} &=& A_1 \oplus A_2 \oplus B_1  \oplus B_2 . \nonumber
\end{eqnarray}
Similarly, we can obtain the basis polynomials
for $C_{2v}$ representation, which are listed
in Table~\ref{table:basis:c2v} for $l\le 2$.
\begin{table}[h]
\begin{ruledtabular}
\begin{tabular}{c|lll}
\mbox{representation}  &  $l=0$  &  $l=1$  &  $l=2$ \\
\hline
$A_1$ & $1$  & $x_3$  & $x_3^2 - \frac{1}{3}x^2$ \\
$A_2$ &      &        & $x_1 x_2$                \\
$B_1$ &      & $x_1$  & $x_1 x_3$                \\
$B_2$ &      & $x_2$  & $x_2 x_3$                \\
\end{tabular}
\end{ruledtabular}
\caption{\label{table:basis:c2v}
The basis polynomials for the irreducible representations of $C_{2v}$.
}
\end{table}

In a typical lattice calculation,
we can easily investigate is the $A_1$ sector.
We will therefore concentrate on this sector.
As is seen,  up to $l\leq 2$, $s$-wave, $p$-wave and $d$-wave
contribute to this sector.
The other symmetry sectors can be easily worked out in the same way.

First, let us consider the angular momentum cutoff $\Lambda=0$.
From the reduction relations~(\ref{eq:decomp_c4v},\ref{eq:decomp_c2v})
and Tables~\ref{table:basis:c4v},~\ref{table:basis:c2v},
we note that only $\mathcal{M}^{\mathbf d}_{00,00}$ belongs to
this sector, and Eq.~(\ref{determinant}) can be written as
\begin{equation}
\label{det-1dim}
\tan\delta_0(k) = \frac{1}{{\mathcal M}^{\mathbf d}_{00,00}} =
\frac{\gamma q \pi^{3/2}}{{\mathcal Z}^{\mathbf d}_{00}(1;q^2)} ,
\qquad  q = \frac{L}{2\pi}k .
\end{equation}
This is our essential result.

If the angular momentum cutoff $\Lambda=1$,
we should  take the sector $l=1$ into consideration,
and the matrix in Eq.~(\ref{determinant}) is two-dimensional.
Hence, the determinant condition contains
both phase shifts $\delta_0$ and $\delta_1$,
\begin{eqnarray}
\label{det-2dim}
&&\hspace{-0.8cm}\left[e^{2i\delta_0}(m_{00}-i) - (m_{00}+i)\right]
\left[e^{2i\delta_1}(m_{11}-i) - (m_{11}+i)\right] \cr
&=&
m_{10}^2 \left(e^{2i\delta_0} - 1\right)\left(e^{2i\delta_1} - 1\right) ,
\end{eqnarray}
where we denote $m_{ab} \equiv {\mathcal M}^{\mathbf d}_{a0,b0}$.
If $\delta_1=0$ (mod $\pi$),
as what we expected, Eq.~(\ref{det-2dim}) reduces nicely
to Eq.~(\ref{det-1dim}).
If $\delta_1 \ne 0$.
Normally we can reasonably suppose that
the lowest $l$-channel  dominate the scattering phase.
This is quite right in the low-energy elastic scattering~\cite{Luscher:1991cf,Rummukainen:1995vs}.

If we expand $\delta_0 = \delta_0^0 + \Delta_0$,
where $\delta_0^0$ meets Eq.~(\ref{det-1dim})
and $\Delta_0$ is a perturbative term,
we can give the first order correction, namely,
$$
\Delta_0(k) = \sigma(k) \delta_1(k) .
$$
The function $\sigma(k)$ stands for the sensitivity of
the higher scattering phases. For $C_{4v}$ symmetry,
it is given by
$$
\sigma(k) = -\frac{m_{10}^2}{m_{00}^2 + 1} ,
$$
which is not usually small.
To make sure that the Eq.~(\ref{det-1dim}) is a well approximation,
the phase shift $\delta_1(k) ({\rm mod}  \pi) $ should be small.
Fortunately, the physical case is often like this way:
the elastic scattering is often dominated
by the lowest $l$- channel.

\begin{figure}[htb]
\begin{center}
\includegraphics[width=8.0cm]{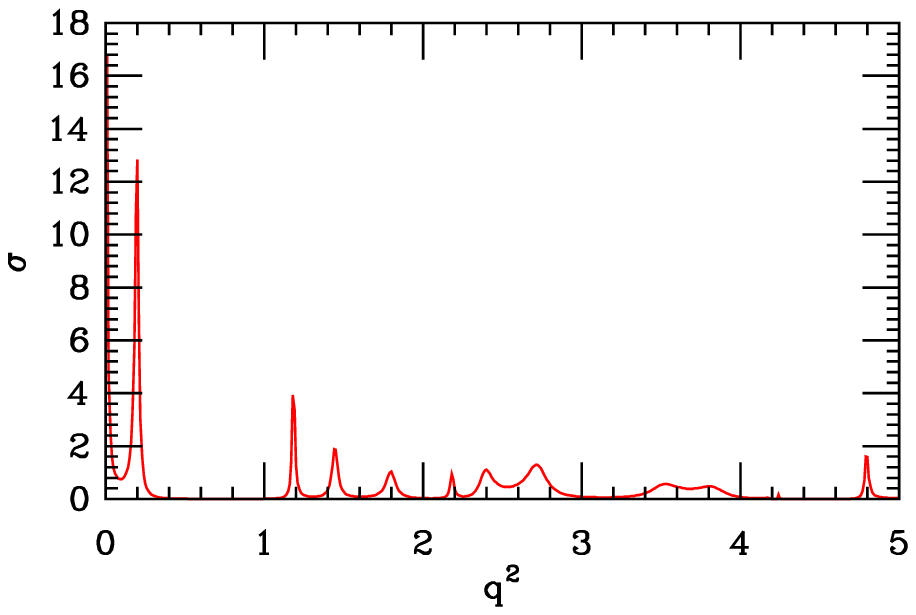}
\includegraphics[width=8.0cm]{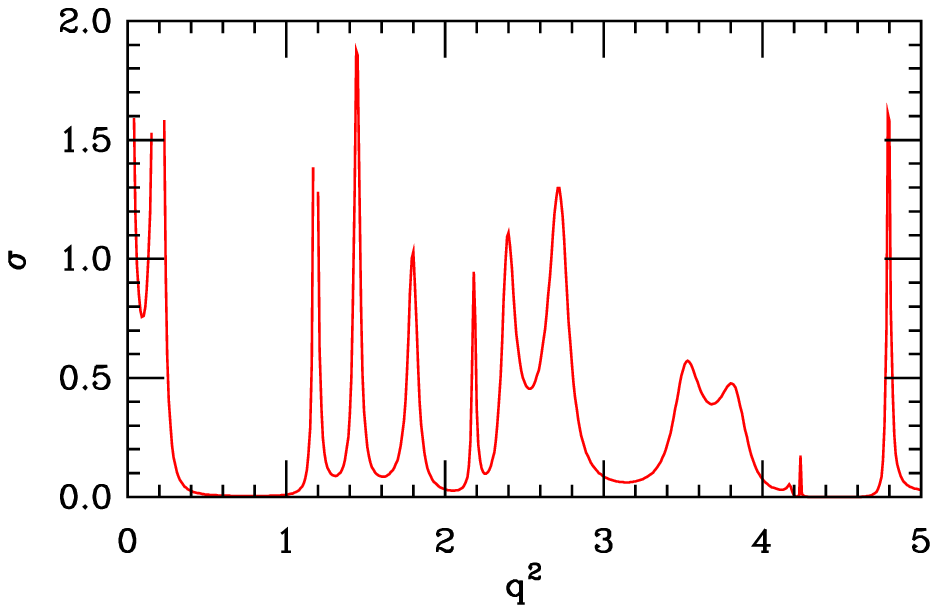}
\end{center}
\caption{\label{fig:sigma}
The sensitivity $\sigma(q^2)$
for $C_{4v}$  symmetry with parameters $\alpha=1.15$ and $\gamma=1.177$.
}
\end{figure}

The sensitivity function $\sigma(q^2)$ can be calculated
using the matrix elements provided in Eq.~(\ref{eq:wlm})
discussed in Appendices~\ref{app:zeta_lm} and \ref{appe:zeta10}.
The sensitivity function $\sigma(q^2)$ for
$C_{4v}$ symmetry with $\alpha=1.15$ and $\gamma=1.177$
is illustrated in Fig.~\ref{fig:sigma},
here $\gamma$ is a boost factor, and
$\alpha$-factor is defined in Eq.~(\ref{alpha_factor}),
which are typical values we used in Ref.~\cite{Fu:2011xw}.
In Fig.~\ref{fig:sigma}
the lower panel  is the same sensitivity function as the upper panel
just to display  some detailed variations.

\begin{figure}[htb]
\begin{center}
\includegraphics[width=8.0cm]{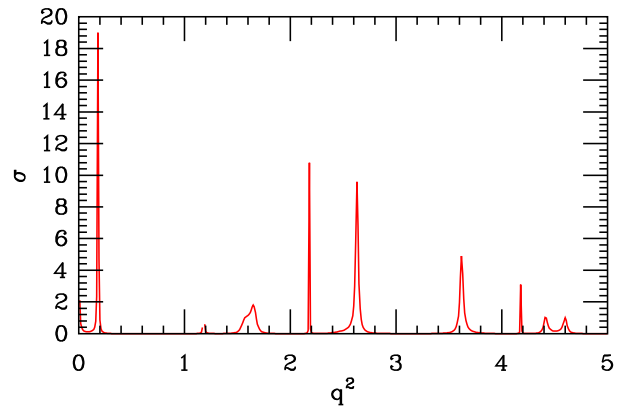}
\end{center}
\caption{\label{fig:sigma_105}
The sensitivity $\sigma(q^2)$
for $C_{4v}$  symmetry with parameters $\alpha=1.05$ and $\gamma=1.177$.
}
\end{figure}

\begin{figure}[htb]
\begin{center}
\includegraphics[width=8.0cm]{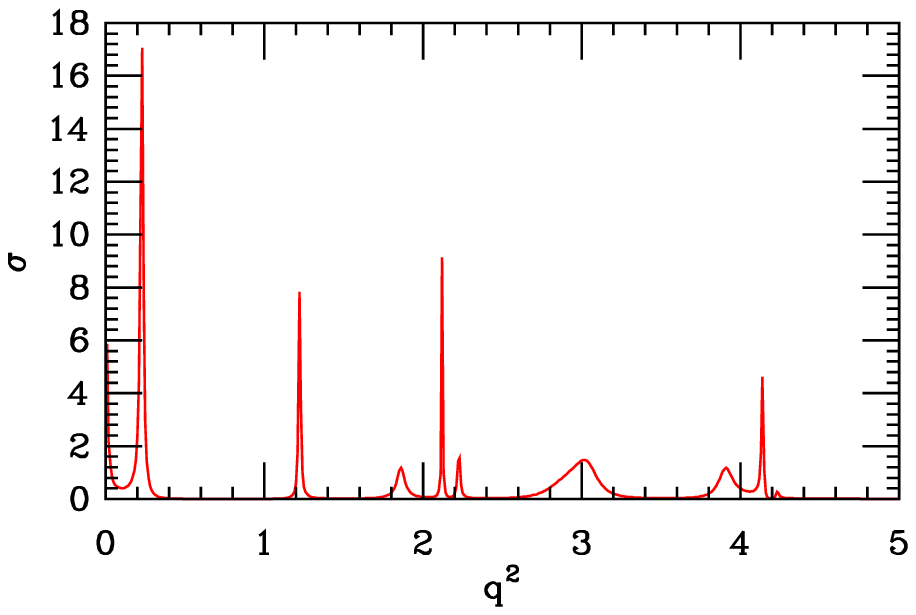}
\end{center}
\caption{\label{fig:sigma_g1_1}
The sensitivity $\sigma(q^2)$
for $C_{4v}$  symmetry with parameters $\alpha=1.1$ and $\gamma=1.067$.
}
\end{figure}

\begin{figure}[htb]
\begin{center}
\includegraphics[width=8.0cm]{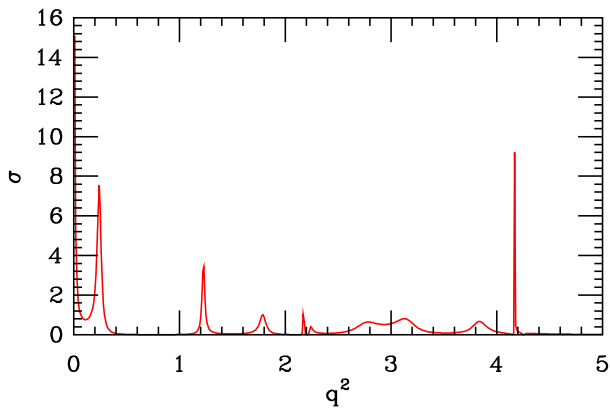}
\end{center}
\caption{\label{fig:sigma_g1_15}
The sensitivity $\sigma(q^2)$
for $C_{4v}$  symmetry with parameters $\alpha=1.15$ and $\gamma=1.067$.
}
\end{figure}

In the work, we also calculate the sensitivity function $\sigma(q^2)$
using the some typical $\alpha$ and $\gamma$ values,
we found that it often varies in the range $0 - 20$,
in Figs.~\ref{fig:sigma_105}, \ref{fig:sigma_g1_1}, and \ref{fig:sigma_g1_15},
we plotted just three of them.
We can note that the sensitivity function $\sigma(q^2)$
is finite for all $q^2 >0$.
For some special values of $q^2$, however,
the sensitivity function $\sigma(q^2)$ has a sharp peak.
For other values of $q^2$ away from these values,
$\sigma(p^2)$ is always not too large.

We also note that, when $q\to 0$,
the sensitivity function $\sigma(q^2)$ is often large.
However, it does not usually cause any problem
because it is nicely canceled out by $\delta_1$
which is of $q^3$ order at small $q$~\cite{Luscher:1991cf,Rummukainen:1995vs}.
Thus, for the range $0 < q^2 < 1.1$ (except some special $q^2$ values),
the Eq.~(\ref{det-1dim}) can be considered a good approximation.
In fact, it is the range which are usually used to
study the elastic scattering from lattice QCD~\cite{Fu:2011xw}.

We should bear in mind that if $\delta_1(k)$ is not small, it is
very difficult to extract the phase shift from energy spectrum~\cite{Rummukainen:1995vs}.
In principle, we can still extract the $s$-wave scattering phase
from Eq.~(\ref{det-2dim}) through dividing the $p$-wave phase shift
by lattice simulations at various energy~\cite{Rummukainen:1995vs},
because the corrections due to higher scattering phases
can also be estimated from lattice calculations.
For example, from Table~\ref{table:basis:c4v},
it is easily noted that, for lattices with $C_{4v}$ symmetry,
by inspecting energy eigenvalues of $E$ symmetry,
we can get an approximate estimate for the $p$-wave scattering phase $\delta_1$
which dominates this symmetry sector.
It seems to be too difficult,  however naturally,
it is still possible to calculate the energy spectrum,
this is our future tasks.

If we choose the sector ${\mathbf d}=0$,
the moving and the CM frames coincide,
$\gamma\rightarrow 1$ and $P_{\mathbf d} \rightarrow {\mathbb Z}^3$,
and Eq.~(\ref{det-2dim}) nicely reduces to
the form given in Ref.~\cite{Luscher:1991cf}.
Of course, if we select $m_1 = m_2$ and
$P_{\mathbf d}\rightarrow \left\{
{\mathbf r}\in{\bf R}^3\,\left|\,{\mathbf r} = \vec\gamma^{-1}
\left({\mathbf n} + {\mathbf d}/2 \right) \right.
\right\}, \; {\mathbf n} \in {\mathbb Z}^3$,
and Eq.~(\ref{det-2dim})  neatly reduces to the form
presented in Ref.~\cite{Rummukainen:1995vs}.
These are what we expected.

As for $\Lambda=2$ or higher, it is quite complicated.
See the relevant discussions in Ref.~\cite{Feng:2004ua}.
Bearing in mind that this work is an exploratory study
for some systems like the $\pi K$ system,
the main purpose is to present some conceptual and theoretical issues.

\section{Conclusion}
\label{sec:Conclusions}
In the current work we have strictly investigated
the scattering states of two-particle with unequal mass,
and the best-efforts are paid to derive the modified $\mathbf d$-periodic rule
which is crucial to the alteration of the Rummukainen-Gottlieb's formula.
The finite size expressions,
which can be regarded as a generalization of the
Rummukainen-Gottlieb's formulae to the generic two-particle system
in the moving frame, are developed.
We also checked that all the Rummukainen-Gottlieb's results
in Ref.~\cite{Rummukainen:1995vs} is nicely restored if we set $m_1=m_2$.

Since the so-called $\kappa$ meson is a low-lying scalar meson
with strangeness, a study of $\kappa$  meson decay is an
explicit exploration of the three-flavor structure of the
low-energy hadronic interactions, which is not directly
probed in $\pi\pi$ scattering, therefore, it is a significant step
for us understanding the dynamical aspect of hadron reactions with QCD.
Moreover, BES collaboration recently carried out
some experimental measurements~\cite{Ablikim:2010kd,Ablikim:2005ni}
to investigate $\kappa$ resonance mass and its decay width.
With the modified formula in Eq.~(\ref{det-1dim})
and our strict discussion of this formula from theoretical aspects,
now it will be possible to compute the resonance masses
and perhaps its decay widths of some resonances
including possible exotic hadrons as well as traditional
hadrons like $\kappa$ and vector kaon $K^*$, etc., directly
from lattice simulation in a correct manner.
We have already used these formulae to preliminarily analyze
our $\pi K$ scattering at $I=1/2$ channel~\cite{Fu:2011xw},
and the reasonable results of our lattice simulation data
supports these formula.

\section*{Acknowledgments}
The author thanks Naruhito Ishizuka
for kindly helping us about group symmetry,
and we would also like to thank Sasa Prelovsek,
Carleton DeTar, and Martin J. Savage
for their encouraging and enlightening comments.

\appendix
\section{The calculation of zeta function
\label{app:zeta_lm}}
The method for evaluating the zeta function for ${\mathbf d=0}$ has been
discussed by L\"uscher in Ref.~\cite{Luscher:1991cf}.
Rummukainen and Gottlieb  extended this discussion
in the MF for ${\mathbf d \ne 0}, \alpha = 1$~\cite{Rummukainen:1995vs}.
The formalism used here is further adapted to
the case of ${\mathbf d}\ne 0, \alpha \ne 1$,
and we just present the essential formulae.

We first denotes the heat kernel
on a modified ${\mathbf d}$-periodic torus in Eq.~(\ref{cm_period}), namely,
\begin{equation}
\label{eq_ker}
K_{\mathbf d}(t, {\mathbf x}) =
\frac{1}{(2\pi)^{3}} \sum_{ {\mathbf r} \in P_{\mathbf d} }
e^{i{\mathbf r}\cdot{\mathbf x} - t{\mathbf r}^2}
\end{equation}
where the summation for ${\mathbf r}$ is carried out over the set
\begin{equation}
\label{eq:app:pset}
P_{\mathbf d} = \left\{ {\mathbf r} \left|  {\mathbf r} =
\vec{\gamma}^{-1}
\left({\mathbf n}+\frac{\alpha}{2} {\mathbf d} \right) \,, \quad
{\mathbf n}\in \mathbb{Z}^3 \right. \right\} ,
\end{equation}
here the factor $\alpha$ is denoted in Eq.~(\ref{alpha_factor}),
and the operation $\vec{ \gamma }^{ -1 }$ is
is defined in Eq.~(\ref{short_gamma}).
Following from Poisson's identity, we can rewrite the heat kernel as
\begin{eqnarray}
\hspace{-0.8cm} K_{\mathbf d}(t, {\mathbf x}) &=&
\gamma \frac{1}{(4\pi t)^{\frac{3}{2}}}
e^{i 1/2 \alpha  {\mathbf d} \cdot {\mathbf x}} \cr
&\times& \sum_{{\mathbf n}\in {\mathbb Z}^3}
e^{-i\alpha \pi {\mathbf d} \cdot {\mathbf n}}
\exp\left[-\frac{1}{4t}({\mathbf x}-2\pi\vec\gamma\,{\mathbf n})^2\right] \,.
\label{kernel2}
\end{eqnarray}
The expression in Eq.~(\ref{eq_ker})
is fast convergent for large $t$,
and the expression in Eq.~(\ref{kernel2}) is useful
for small $t$.  We can denote the truncated heat kernel
$K_{{\mathbf d}}^{\lambda}(t, {\mathbf x})$ by
$$
K_{{\mathbf d}}^{\lambda}(t, {\mathbf x}) =
K_{\mathbf d}(t,{\mathbf x}) -
\sum_{ {\mathbf r} \in P_{\mathbf d}, {|{\mathbf r}| < \lambda} }
\exp(i{\mathbf r}\cdot{\mathbf x} - t{\mathbf r}^2) .
$$
We apply the operator ${\mathcal Y}_{lm}(-i\nabla_{\mathbf x})$ to
heat kernels,
\begin{eqnarray}
{\mathcal K}^\lambda_{{\mathbf d},lm}(t,{\mathbf x})
&=&
{\mathcal Y}_{lm}(-i\nabla_{\mathbf x})
{\mathcal K}_{{\mathbf d}}^\lambda(t,{\mathbf x}) \;.
\end{eqnarray}
We can easily show that the zeta function has a rapidly convergent
integral expression
\begin{eqnarray}
\hspace{-0.6cm}{\mathcal Z}^{\mathbf d}_{lm}(1;q^2)
\hspace{-0.2cm}&=&\hspace{-0.2cm}
\sum_{ {\mathbf r} \in P_{\mathbf d}, {|{\mathbf r}| < \lambda} }
\frac{{\cal Y}_{lm}({\mathbf r})}{{\mathbf r}^2 - q^2}
\cr
\hspace{-0.2cm}&+&\hspace{-0.2cm}
(2\pi)^3 \hspace{-0.2cm} \int_0^\infty \hspace{-0.3cm} dt \left( e^{tq^2}
K_{{\mathbf d},lm}^{\lambda}(t,{\mathbf 0}) \hspace{-0.1cm}-\hspace{-0.1cm}
\frac{\gamma\delta_{l,0}\delta_{m,0}}{16\pi^2 t^{3/2}}\right).
\end{eqnarray}
To calculate the integrand, we use the Eq.~(\ref{eq_ker})
when $t \ge 1$, and  the Eq.~(\ref{kernel2})
in the case of $t < 1$.
The cutoff $\lambda$ is chosen such
that $\lambda^2 > \mbox{Re\,} q^2$.
We can easily verify that, when $m_1=m_2$ (or equivalently $\alpha=1$),
the Rummukainen-Gottlieb's result in Ref.~\cite{Rummukainen:1995vs}
is restored.

\section{The evaluation of the zeta function $\mathcal{Z}_{10}(s;q^2)$}
\label{appe:zeta10}
In this appendix we briefly discuss one useful method
for numerical evaluation of zeta function $\mathcal{Z}_{10}(s;q^2)$.
Here we follow the methods and notations in Ref.~\cite{Yamazaki:2004qb}.

The definition of the zeta function $\mathcal{Z}_{10}^{\mathbf d}(s;q^2)$
in Eq.~(\ref{zeta_Func}) is
\begin{equation}
\sqrt{\frac{4\pi}{3} } \cdot \mathcal{Z}^{ \mathbf d }_{ 10 } ( s ; q^2 ) =
\sum_{ {\mathbf r} \in P_{\mathbf d} } \frac{ r_3 }{( r^2 - q^2 )^s } \,,
\label{eq:Z00d_appendix}
\end{equation}
where the summation for ${\mathbf r}$ is carried out over the set
\begin{equation}
\label{eq:app:p_set2}
P_{\mathbf d} = \left\{ {\mathbf r} \left|  {\mathbf r} =
\vec{\gamma}^{-1}
\left({\mathbf n}+\frac{\alpha}{2} {\mathbf d} \right), \quad
{\mathbf n}\in \mathbb{Z}^3 \right. \right\} ,
\end{equation}
here the factor $\alpha$ is denoted in Eq.~(\ref{alpha_factor}).
The operation $\vec{ \gamma }^{ -1 }$ is
is defined in Eq.~(\ref{short_gamma}).
We consider that the value $q^2$ can be a positive or negative.

First we consider the case of $q^2 > 0$,
and we can separate the summation in $\mathcal{Z}_{10}^{\mathbf d}( s ; q^2 )$
into two parts as
\begin{equation}
\sum_{{\mathbf r} \in P_{\mathbf d}} \frac{r_3}{( r^2 - q^2 )^{s}} =
\sum_{r^2 < q^2}\frac{r_3}{(r^2-q^2)^{s}} +
\sum_{r^2 > q^2}\frac{r_3}{(r^2-q^2)^{s}} \,.
\label{eq:app_zeta_1}
\end{equation}
The second term can be written in an integral form,
\begin{widetext}
\begin{eqnarray}
\sum_{ r^2 > q^2 } \frac{r_3}{( r^2 - q^2 )^{s}}
&=&
\frac{1}{ \Gamma(s) } \sum_{r^2 > q^2 } r_3
\left[ \int_0^1        {\rm d}t \ t^{s-1}  e^{ - t ( r^2 - q^2) } +
       \int_1^{\infty} {\rm d}t \ t^{s-1} e^{ - t ( r^2 - q^2) } \right] \cr
&=&
\frac{1}{\Gamma(s)}
\int_0^1 {\rm d}t  t^{s-1} e^{q^2 t} \sum_{\mathbf r} r_3 e^{-r^2 t}-
\sum_{ r^2 < q^2 } \frac{r_3}{( r^2 - q^2 )^{s}} +
\sum_{\mathbf r} r_3 \frac{ e^{ -( r^2 - q^2 ) } }{ (r^2 - q^2)^s} \,.
\label{eq:app_zeta_2}
\end{eqnarray}
The second term neatly cancel out the first term in Eq.~(\ref{eq:app_zeta_1}).
Using Poisson's resummation formula
we can rewrite the first term in Eq.~(\ref{eq:app_zeta_2}) as
\begin{eqnarray}
\frac{1}{\Gamma(s)} \int_0^1  {\rm d}t  t^{s-1} e^{ t q^2 }
\sum_{\mathbf r } r_3 e^{-r^2 t} &=&
\frac{\sqrt{\pi} }{ \Gamma(s) }
\int_0^1  {\rm d}t  t^{s-1} e^{t q^2}
\left( \frac{ \pi }{t} \right)^{2}
\sum_{ {\mathbf n } \in \mathbb{Z}^3 } i \, n_3
e^{ - i \alpha \pi \mathbf{n} \cdot \mathbf{d}  }
e^{ - ( \pi \vec{\gamma}{\mathbf n} )^2/t } \cr
&=&
\frac{\sqrt{\pi} }{ \Gamma(s) }
\int_0^1  {\rm d}t  t^{s-1} e^{t q^2}
 \left( \frac{ \pi }{t} \right)^{2}
\sum_{ {\mathbf n } \in \mathbb{Z}^3 } n_3
\sin( \alpha \pi \mathbf{n} \cdot \mathbf{d} )
e^{ - ( \pi \vec{\gamma}{\mathbf n} )^2/t } \,,
\label{eq:app_zeta_3}
\end{eqnarray}
\end{widetext}
where the imaginary parts are neatly canceled out.

After gathering all terms we obtain
the zeta function  at $s=1$ as,
\begin{eqnarray}
\sqrt{\frac{4\pi}{3} } \cdot \mathcal{Z}_{10}^{\mathbf d} (1; q^2) &=&
\sum_{ {\mathbf r} \in P_{\mathbf d} }
r_3 \frac{ e^{ -(r^2-q^2) } }{r^2-q^2} + \cr
&&\hspace{-4.0cm}\sqrt{\pi} \int_0^1  {\rm d}t  e^{t q^2}
 \left( \frac{ \pi }{t} \right)^{2}
\sum_{ {\mathbf n } \in \mathbb{Z}^3 } n_3
\sin( \alpha \pi {\mathbf n} \cdot {\mathbf d} )
e^{ - (\pi \vec{\gamma}{\mathbf n} )^2/t } .
\label{eq:app_zeta_s=1}
\end{eqnarray}

For the case of $q^2 \le 0$, it is not necessary for us to separate the summation
in $\mathcal{Z}_{10}(s; q^2)$, and it can be also written in an integral form.
Following the same procedures, we arrive at the same expression
in Eq.~(\ref{eq:app_zeta_s=1}).
Hence, Eq.~(\ref{eq:app_zeta_s=1}) can be applied for both cases.

Substituting ${\mathbf d} = (0,0,1)$ into Eq.~(\ref{eq:app_zeta_s=1})
we obtain the zeta function in Eq.~(\ref{zeta_Func})
\begin{eqnarray}
&&\hspace{-1.3cm}\sqrt{\frac{4\pi}{3} } \cdot
\mathcal{Z}_{ 10 }^{\mathbf d} (1; q^2) =
\sum_{{\mathbf r} \in P_{\mathbf d}} r_3 \frac{ e^{ -(r^2-q^2) } }{r^2-q^2}
+ \cr
&&\hspace{-1.2cm}
\sqrt{\pi} \int_0^1  {\rm d}t  \, e^{t q^2}
\left( \frac{ \pi }{t} \right)^{2}
\sum_{ {\mathbf n } \in \mathbb{Z}^3 } n_3
\sin( \alpha  \pi n_3 )
e^{ - ( \pi \vec{\gamma}{\mathbf n} )^2/t } \, .
\end{eqnarray}

We can easily verify that, if $m_1=m_2$,
zeta function $\mathcal{Z}_{10}(1; q^2 ) \to 0 $,
the Rummukainen-Gottlieb's result is recovered.


\end{document}